\documentclass[11pt]{article}
\usepackage[a4paper]{geometry}
\usepackage{amsmath}
\usepackage{amsfonts}
\usepackage{amsbsy}
\usepackage{amsthm}
\usepackage{accents}
\usepackage{tensor}
\usepackage{authblk}
\usepackage{braket}
\usepackage{graphicx}
\usepackage{caption}
\usepackage{subcaption}
\usepackage{cite}

\newcommand{\bs}[1]{\mathbf{#1}}
\newcommand{\bsv}[1]{\boldsymbol{#1}}
\newcommand{\kn}{|\vec{k}|}
\newcommand{\ofk}{(\vec{k})}
\newcommand{\volel}{\frac{d^3 k}{(2 \pi)^3 \kn}}
\newcommand{\rtt}{\widetilde{\mathbb{R}}^3}
\newcommand{\dcon}{\hat{\vec{D}}}
\newcommand{\tPsi}{\widetilde{\Psi}}
\newcommand{\tPhi}{\widetilde{\Phi}}
\newcommand{\dcp}{\hat{D}'}
\newcommand{\gmc}[1]{\gamma_{#1} \big[ C \big]}
\newcommand{\hrm}{\dag}

\DeclareMathOperator{\Tr}{Tr}

\theoremstyle{definition}
\newtheorem{exmpl}{Example}
\newtheorem{rmrk}{Remark}

\numberwithin{equation}{section}

\begin{document}
\title{\bf The geometrical interpretation of the photon position operator} 
\date{\today} 
\author[1]{Michał~Dobrski\thanks{michal.dobrski@p.lodz.pl}}
\author[1]{Maciej~Przanowski\thanks{Professor emeritus}\thanks{maciej.przanowski@p.lodz.pl}}
\author[1]{Jaromir~Tosiek\thanks{jaromir.tosiek@p.lodz.pl}}
\author[2]{Francisco~J.~Turrubiates\thanks{fturrubiatess@ipn.mx}}
\affil[1]{Institute of Physics, Łódź University of Technology, Wólczańska 219, 90-924 Łódź, Poland}
\affil[2]{Departamento de Física, Escuela Superior de Física y Matemáticas, Instituto~Politécnico~Nacional, Unidad Adolfo López Mateos, Edificio 9, 07738~Ciudad~de~México, México} 
\maketitle 
\begin{abstract}
It is shown that the photon position operator $\hat{\vec{X}}$ with commuting components can be written in the momentum representation as $\hat{\vec{X}}=i \dcon$, where $\dcon$ is a flat connection in the tangent bundle $T(\mathbb{R}^3 \setminus \{ (0,0,k_3) \in \mathbb{R}^3 : k_3 \geq 0\})$ over $\mathbb{R}^3 \setminus \{ (0,0,k_3) \in \mathbb{R}^3 : k_3 \geq 0\}$ equipped with the Cartesian structure. Moreover, $\dcon$ is such that the tangent $2$-planes orthogonal to the momentum are parallelly propagated with respect to $\dcon$ and, also, $\dcon$ is an anti-Hermitian operator with respect to the scalar product $\langle \bs{\Psi} | \hat{H}^{-2s} |\bs{\Phi} \rangle$. The eigenfunctions $\bs{\Psi}_{\vec{X}} (\vec{x})$ of the position operator $\hat{\vec{X}}$ are found.
\end{abstract}

\section{Introduction}
T.~D.~Newton and E.~P.~Wigner in their distinguished paper on localized states of relativistic quantum particles \cite{newton}, among others, came to the result that for massless particles with the spin~$\geq 1$ no localized states in the sense explained in the paper exist. Then, the authors conclude: ``This is an unsatisfactory, if not unexpected, feature of our work''. A.~S.~Wightman \cite{wightman} also found that the photon is not localizable. In contrary using some slightly weaker requests for localizability, one can show \cite{jauch, amrein} that the photon is localizable. M.~H.~L.~Pryce in his pioneering work devoted to the mass-centre in relativistic field theory \cite{pryce} has found an operator which then he considers as the photon position operator. It is Hermitian operator but unfortunately its components do not commute. In 1999 Margaret Hawton introduced the photon position operator with commuting components \cite{hawton}. She demonstrated that this new operator differs from the Pryce operator only by one term, which turned out to be closely related to the Berry potential leading to the Berry phase whose appearance has been proved experimentally \cite{chiao,tomita} and discussed in detail from the group theoretical and geometrical points of view by I.~Białynicki-Birula and Z.~Białynicka-Birula in \cite{bialynicki}. Hawton's position operator for the photon has then been widely investigated in the next works \cite{hawton2,hawton3,hawton4,hawton5,debierre,hawton6}. Our work follows this path. In Section 2 we present a simple derivation of the Hawton position operator for the photon. We will show next that this operator in momentum representation, in terms of differential geometry, can be interpreted as some covariant derivative (connection) multiplied by imaginary unit $i$. This covariant derivative is defined in the Cartesian tangent bundle $T(M)$, where $M$ is a dense open submanifold of the Cartesian momentum space $\mathbb{R}^3$; the respective curvature tensor is zero (the connection is flat), $2$-planes orthogonal to the momentum are parallely propagated and, finally, the covariant derivative is an anti-Hermitian operator with respect to a given scalar product. We were able to find the general operator arising in this way. Then we find a phase space image of 
the photon position operator in the sense of the Weyl-Wigner-Moyal formalism. In Section 3 we find the eigenfunctions $\bs{\tPsi}_{\vec{X}} \ofk$ of the photon position operators in the momentum representation and then we calculate the inverse Fourier transforms of those eigenfunctions which give us the wave eigenfunctions $\bs{\Psi}_{\vec{X}} (\vec{x})$. It is demonstrated that $\bs{\Psi}_{\vec{X}} (\vec{x})$ is localized in a small neighborhood of $\vec{x} = \vec{X}$. This result is compatible with the interpretation of the photon wave function $\bs{\Psi} (\vec{x})$ given in \cite{bialynicki2,bialynicki3,sipe}. Section 4 is devoted to analysis of the Berry potential and the Berry phase related to the Hawton's position operator. 
Although several results of our paper have been found previously by other authors, we hope that the geometrical interpretation given here provides a new perspective on the photon position operator.

\section{Relation between Hawton's position operator of the photon and a covariant derivative (connection)}
In this section we show that the photon position operator introduced by Margaret Hawton \cite{hawton} is closely related to some covariant derivative (connection) in the tangent bundle over the differential manifold $\mathbb{R}^3 \setminus \{ (0,0,c) \in \mathbb{R}^3 : c \geq 0\}$ with the Cartesian structure. In our work we adopt quantum mechanics of the photon as developed by I.~Białynicki-Birula \cite{bialynicki2,bialynicki3} and J.~E.~Sipe \cite{sipe}, and reconstructed in \cite{przanowski}. Using the notation of \cite{przanowski} one concludes that the photon wave function in the Cartesian momentum coordinates is represented by the complex vector function 
\begin{equation}
\bs{\tPsi} \ofk = 
\begin{pmatrix}
\tPsi_1 \ofk \\
\tPsi_2 \ofk \\
\tPsi_3 \ofk 
\end{pmatrix},
\quad \quad \vec{k}=(k_1,k_2,k_3)\in \mathbb{R}^3
\end{equation}
perpendicular to the vector $\vec{k}$
\begin{equation}
\label{ortcond}
k_j \tPsi_j \ofk = 0
\end{equation}
(summation over $j$ from $1$ to $3$).

Since the components of the metric tensor in momentum space are given by the Kronecker delta $\delta_{jl}$, there is no difference between covariant and contravariant components in the Cartesian coordinates $(k_1,k_2,k_3)$. We are looking for the photon position operator in momentum representation. In standard quantum mechanics we have
\begin{equation}
\label{stdop}
\hat{\vec{x}} = i \vec{\nabla}, \quad \quad \vec{\nabla} \equiv \left( \frac{\partial}{\partial k_1}, \frac{\partial}{\partial k_2}, \frac{\partial}{\partial k_3}  \right) = \left( \partial_1, \partial_2, \partial_3  \right)
\end{equation}
However one quickly notes that with (\ref{ortcond}) satisfied 
\begin{align}
& & k_j \left( \hat{x}_l \bs{\tPsi} \ofk \right)_j &= -i \tPsi_j \ofk \partial_l k_j = - \tPsi_l \ofk, & l&=1,2,3
\end{align}
Therefore $\hat{\vec{x}} \bs{\tPsi} \ofk $ does not fulfill the condition (\ref{ortcond}) for $\bs{\tPsi} \neq 0$ although $\bs{\tPsi} \ofk $ does. This means that the standard position operator (\ref{stdop}) is definitely not the photon position operator and, consequently, a generalization of (\ref{stdop}) is needed. We propose to assume that the position operator of the photon $\hat{\vec{X}}$ has the form 
\begin{align}
\label{posopdef}
& & \hat{\vec{X}} &= i \dcon \Longrightarrow  \hat{X}_l = i \hat{D}_l, & l&=1,2,3
\end{align}
where $\dcon=(\hat{D}_1,\hat{D}_2,\hat{D}_3)$ is an operator in the Hilbert space $L^2(\mathbb{R}^3) \otimes \mathbb{C}^3$ determined by a suitable \emph{covariant derivate (the connection)} in the tangent bundle $T(M)$ over $M \subset \mathbb{R}^3$, with $M$ being some dense open submanifold of $\mathbb{R}^3$ endowed with the Cartesian structure. This connection we also denote by $\dcon$. Since we require the components $\hat{X}_1, \hat{X}_2, \hat{X}_3$ of $\hat{\vec{X}}$ to commute, we must assume that the operators $(\hat{D}_1, \hat{D}_2, \hat{D}_3)$ mutually commute
\begin{align}
\label{flat1}
& & \left[ \hat{D}_l, \hat{D}_m \right]&=0, & l,m&=1,2,3
\end{align}
This last condition implies the vanishing of the curvature of the connection $\dcon$
\begin{align}
\label{flat2}
& & \tensor{R}{^j_{lmn}}&=0, & j,l,m,n&=1,2,3
\end{align}
It means that the connection $\dcon$ is flat \cite{kobayashi,kobayashi2}. Then one expects that the operators $\hat{X}_l$, $l=1,2,3$, acting on a photon wave function $\bs{\tPsi} \ofk$ give also a photon wave function. This assumption implies that the connection $\dcon$ has to fulfill the following conditions
\begin{align}
\label{connort}
& & k_j \left( \hat{D}_l \bs{\tPsi} \ofk \right)_j&=0,& l&=1,2,3
\end{align}
for any section $\bs{\tPsi} \ofk$ of the complexified tangent bundle $T_\mathbb{C} (M)$ satisfying the relation (\ref{ortcond}).

Last but not least, restriction imposed on the connection $\dcon$ follows from the fact that the photon position operator $\hat{\vec{X}}$ should be Hermitian operator with respect to the \emph{Białynicki-Birula scalar product} \cite{bialynicki3,hawton,hawton2,przanowski}
\begin{equation}
\label{bbsp}
\langle \bs{\tPhi} | \bs{\tPsi} \rangle_{\mathrm{BB}} := \langle \bs{\tPhi} | \hat{H}^{-1} |\bs{\tPsi} \rangle = \int \volel \bs{\tPhi}^{\hrm} \ofk \bs{\tPsi} \ofk
\end{equation}
where $\hat{H}$ is the photon Hamiltonian operator which in momentum representation reads
\begin{equation}
\hat{H}=c \hbar \kn
\end{equation}
Consequently, $\hat{X}_l$ for $l=1,2,3$ has to satisfy the relations
\begin{align}
& & \left( \int \volel \bs{\tPhi}^{\hrm} \ofk \hat{X}_l \bs{\tPsi} \ofk \right)^* &= \int \volel \bs{\tPsi}^{\hrm} \ofk \hat{X}_l \bs{\tPhi} \ofk
& l&=1,2,3
\end{align}
for any photon wave functions $\bs{\tPhi} \ofk$ and $\bs{\tPsi} \ofk$. Hence, by (\ref{posopdef}), we assume that the covariant derivative $\dcon$ is an anti-Hermitian operator with respect to the scalar product (\ref{bbsp}). So
\begin{align}
\label{antiherm}
& & \left( \int \volel \bs{\tPhi}^{\hrm} \ofk \hat{D}_l \bs{\tPsi} \ofk \right)^* &= - \int \volel \bs{\tPsi}^{\hrm} \ofk \hat{D}_l \bs{\tPhi} \ofk
& l&=1,2,3
\end{align}
for sections $\bs{\tPhi} \ofk$, $\bs{\tPsi} \ofk$ of the complexified tangent bundle $T_\mathbb{C} (M)$.

Now we are going to construct the covariant derivative (the connection) which satisfies the above conditions (\ref{flat1}) (or, equivalently, (\ref{flat2})), (\ref{connort}) and (\ref{antiherm}). First one should decide the question what the submanifold $M \subset \mathbb{R}^3$ is. We show that $M \neq \mathbb{R}^3$. To this end assume the opposite, that $M = \mathbb{R}^3$. The set of real tangent vectors at the point $(k_1,k_2,k_3) \neq (0,0,0)$ fulfilling the orthogonality condition (\ref{ortcond}) constitutes the $2$-plane $\Pi^{\bot} (k_1,k_2,k_3)$ perpendicular to the vector $\vec{k} = (k_1,k_2,k_3)$. Thus we obtain a $2$-dimensional differential distribution
\begin{equation}
\label{distrib}
\mathcal{D}_2 : \mathbb{R}^3 \ni (k_1,k_2,k_3) \mapsto \Pi^{\bot} (k_1,k_2,k_3)
\end{equation}
Then the condition (\ref{connort}) means that the planes $\Pi^{\bot} (k_1,k_2,k_3)$ are parallelly propagated with respect to the connection $\dcon$ and the condition (\ref{flat1}) (or, equivalently, (\ref{flat2})) says, as has been pointed out above, that $\dcon$ is flat. The integral manifolds of the distribution $\mathcal{D}_2$ are the $2$-spheres  $\kn=\mathrm{const}$. One can easily conclude that our assumptions on the connection $\dcon$ imply among other things, that for any $2$-sphere $\kn = \mathrm{const} > 0$ and for any point $(k_1,k_2,k_3)$ of this sphere and any non-zero vector tangent at $(k_1,k_2,k_3)$ to the sphere one can propagate parallelly this vector with respect to $\dcon$ on the whole sphere thus obtaining nowhere vanishisng tangent vector field on the $2$-sphere. As is well known this is imposible for a topological reason (the Euler characteristic of the $2$-sphere is non-zero \cite{sulanke,kobayashi2}). Consequently, $M$ cannot be considered as equal to $\mathbb{R}^3$. To get $M$ one must remove at least one point from each $2$-sphere $\kn = \mathrm{const} > 0$. We decide to follow this minimal restriction and we remove the north pole for the each $2$-sphere $\kn = \mathrm{const} > 0$ and also point $(0,0,0)$ of $\mathbb{R}^3$. Thus the submanifold $M \subset \mathbb{R}^3$ is assumed as
\begin{equation}
M=\rtt := \mathbb{R}^3 \setminus \{ (0,0,k_3) \in \mathbb{R}^3 : k_3 \geq 0\}
\end{equation}
and we also define the Cartesian structure on $M$. Therefore, in geometrical language, our task is \emph{to find a general flat connection $\dcon$ in the tangent bundle $T(\rtt)$, anti-Hermitian with respect to the scalar product (\ref{bbsp}) and such that $2$-planes of the $2$-dimensional differential distribution $\mathcal{D}_2$ defined by (\ref{distrib}) are parallelly propagated with respect to $\dcon$.} To solve this problem we first choose a basis $(\vec{e}_1,\vec{e}_2,\vec{e}_3)_{\vec{k}^{(0)}}$ of the tangent space $T_{\vec{k}^{(0)}} (\rtt)$ at some point $\vec{k}^{(0)}=(k^{(0)}_1,k^{(0)}_2,k^{(0)}_3) \in \rtt$ so that $\vec{e}_1,\vec{e}_2 \in \Pi^{\bot}$. Since $\rtt$ is simply connected and we assume that the connection $\dcon$ is flat, the basis $(\vec{e}_1,\vec{e}_2,\vec{e}_3)_{\vec{k}^{(0)}}$ can be parallelly propagated with respect to $\dcon$ on the entire $\rtt$ giving a triad of the pointwise independent vector fields $(\vec{e}_1 \ofk,\vec{e}_2 \ofk,\vec{e}_3 \ofk)$ on $\rtt$. Moreover, if $\dcon$ satisfies the condition (\ref{connort}), i.e.\ the planes $\Pi^{\bot} (k_1,k_2,k_3)$ of the distribution $\mathcal{D}_2$ are parallelly propagated with respect to $\dcon$, then
\begin{equation}
\vec{e}_1 \ofk, \vec{e}_2 \ofk \in \Pi^{\bot} \ofk, \quad \quad \forall \, \vec{k} \in \rtt
\end{equation}
Denote 
\begin{align}
& & \vec{e}_{\mu} \ofk \equiv \vec{e}_{\mu} &= e_{\mu j} \frac{\partial}{\partial k_j}, & \mu&=1,2,3
\end{align}
Let $(\vec{e}^{\,1} \ofk, \vec{e}^{\,2} \ofk, \vec{e}^{\,3} \ofk)$ be the triad of $1$-forms dual to $(\vec{e}_1 \ofk, \vec{e}_2 \ofk, \vec{e}_3 \ofk)$, 
\begin{align}
& & \vec{e}^{\, \mu} \ofk \equiv \vec{e}^{\, \mu} &= \tensor{e}{^{\mu}_j} dk_j, & \mu&=1,2,3
\end{align}
Thus one has
\begin{align}
\label{dual}
& & \vec{e}^{\, \mu} (\vec{e}_{\nu}) = \delta^{\mu}_{\nu} &\Longleftrightarrow \tensor{e}{^{\mu}_j} e_{\nu j} = \delta^{\mu}_{\nu}, & \mu, \nu &=1,2,3
\end{align}
From (\ref{dual}) we quickly infer that
\begin{equation}
\label{dual2}
\tensor{e}{^{\mu}_j} e_{\mu l} = \delta_{jl}
\end{equation}
The metric tensor $g_{\mu \nu}$ in the basis $\vec{e}_{\mu}$ reads
\begin{equation}
\label{gdef}
g_{\mu \nu} = \vec{e}_{\mu} \cdot \vec{e}_{\nu} = e_{\mu j} e_{\nu j}
\end{equation}
From (\ref{dual}) and (\ref{gdef}) one easily finds the relations
\begin{align}
\label{erels}
e_{\mu j} &= g_{\mu \nu} \tensor{e}{^{\nu}_j} & \tensor{e}{^{\mu}_j} &= g^{\mu \nu} e_{\nu j}
\end{align}
where, as usual, $g^{\mu \nu}$ is the inverse tensor to $g_{\mu \nu}$
\begin{equation}
g_{\mu \rho} g^{\rho \nu} = \delta^{\mu}_{\nu}
\end{equation}
From the assumption that the vector fields $\vec{e}_{\mu}$, $\mu=1,2,3$ are parallelly propagated with respect to the connection $\dcon$ one has
\begin{equation}
\label{dezero}
\hat{D}_l  \vec{e}_{\mu} = 0 \Longrightarrow \partial_l e_{\mu j} + \Gamma_{jml} e_{\mu m} = 0
\end{equation}
where $\Gamma_{jml}$ are the \emph{connection coefficients}. Multiplying both sides of Eq.\ (\ref{dezero}) by $e_{\mu n}$, summing over $\mu$ from $1$ to $3$, and using (\ref{dual2}) one gets the connection coefficients as 
\begin{equation}
\label{gammadef}
\Gamma_{jnl} = - \tensor{e}{^{\mu}_n} \partial_l e_{\mu j}= e_{\mu j} \partial_l \tensor{e}{^{\mu}_n}
\end{equation}
Gathering, Eq.\ (\ref{gammadef}) gives the general form of the connection coefficients of the flat connection $\dcon$ in $T(\rtt)$ for which the $2$-planes $\Pi^{\bot} \ofk$ are parallelly propagated. Therefore, it remains only to study the condition (\ref{antiherm}). The left side of Eq.\ (\ref{antiherm}) with the use of the well known formula for the covariant derivative
\begin{equation}
\left(\hat{D}_l \bs{\tPsi} \right)_j = \partial_l \tPsi_j + \Gamma_{jml} \tPsi_m
\end{equation}
(and analogous formula for $(\hat{D}_l \bs{\tPhi})_j$), after integrating by parts, and after employing (\ref{gammadef}), (\ref{erels}) and (\ref{dual2}) gives 
\begin{equation}
\begin{split}
\left( \int \volel \bs{\tPhi}^{\hrm} \ofk \hat{D}_l \bs{\tPsi} \ofk \right)^* & = 
- \int \volel \bs{\tPsi}^{\hrm} \ofk \hat{D}_l \bs{\tPhi} \ofk \\
& \quad - \int \volel \tPsi^*_n \tensor{e}{^{\mu}_n} \tensor{e}{^{\nu}_j} \left( \partial_l g_{\mu \nu} + g_{\mu \nu} \kn \partial_l \kn^{-1}\right)\tPhi_j
\end{split}
\end{equation}
Comparing this with the right side of (\ref{antiherm}) one concludes that $\dcon$ is an anti-Hermitian operator with respect to the scalar product (\ref{bbsp}) iff
\begin{equation}
\label{gcond}
\partial_l g_{\mu \nu} +  \left( \kn \partial_l \kn^{-1} \right) g_{\mu \nu} = 0 \Longleftrightarrow \partial_l \left( \kn^{-1} g_{\mu \nu}\right) =0 \Longleftrightarrow  \kn^{-1} g_{\mu \nu} = \mathrm{const}_{\mu \nu}
\end{equation}
Without any loss of generality we can choose the triad $\vec{e}_{\mu} \ofk$ and its dual $\vec{e}^{\,\mu} \ofk$ as (see (\ref{gdef}), (\ref{gammadef}) and (\ref{gcond}))
\begin{equation}
\label{Edef}
\begin{split}
\vec{e}_{\mu} \ofk &= \kn^{\frac{1}{2}} \vec{E}_{\mu} \ofk  \\
\vec{e}^{\, \mu} \ofk &= \kn^{-\frac{1}{2}} \vec{E}^{\,\mu} \ofk, \quad \quad \mu = 1,2,3
\end{split}
\end{equation}
where $(\vec{E}_1, \vec{E}_2, \vec{E}_3)$ is the orthonormal right oriented triad of vector fields on $\rtt$
\begin{equation}
\label{orthonormalt}
\begin{split}
\vec{E}_{\mu} \ofk \cdot \vec{E}_{\nu} \ofk & = \delta_{\mu \nu},\\
\vec{E}_1 \ofk, \vec{E}_2 \ofk &\in \Pi^{\bot} \ofk, \\
\vec{E}_3 \ofk &= \frac{\vec{k}}{\kn},\\
\vec{E}_1 \ofk \times \vec{E}_2 \ofk &= \vec{E}_3 \ofk \\
\end{split}
\end{equation}
and $\vec{E}^{\, \mu} \ofk$ is the triad of $1$-forms dual to $\vec{E}_{\mu} \ofk$. Inserting (\ref{Edef}) into (\ref{gammadef}) we find the connection coefficients
\begin{equation}
\Gamma_{jnl} = -\frac{k_l}{2 \kn^2} \delta_{jn} + E_{\mu j}\partial_l E_{\mu n} = -\frac{k_l}{2 \kn^2} \delta_{jn} - E_{\mu n} \partial_l E_{\mu j}
\end{equation}
(Note that $E_{\mu n } = \tensor{E}{^{\mu}_n}$). Therefore
\begin{equation}
\begin{split}
\left(\hat{D}_l \bs{\tPsi} \right)_j &= \partial_l \tPsi_j - \frac{k_l}{2 \kn^2} \tPsi_j + E_{\mu j}\left(\partial_l E_{\mu n}\right) \tPsi_n \\
&=
\left( 
\delta_{jn} \partial_l  - \delta_{jn} \frac{k_l}{2 \kn^2}  + E_{\mu j}\left(\partial_l E_{\mu n}\right)
\right) \tPsi_n
\end{split}
\end{equation}
Finally, using (\ref{posopdef}) one finds the photon position operator as
\begin{equation}
\label{posop1}
\begin{split}
\hat{\vec{X}} &= (\hat{X}_1,\hat{X}_2,\hat{X}_3)\\
\left(\hat{X}_l \bs{\tPsi} \right)_j &= i \left( 
\delta_{jn} \partial_l  - \delta_{jn} \frac{k_l}{2 \kn^2}  + E_{\mu j}\left(\partial_l E_{\mu n}\right)
\right) \tPsi_n
\end{split}
\end{equation}
which can be rewritten in the matrix form
\begin{equation}
\label{posop2}
\begin{split}
\hat{X}_l \bs{\tPsi} &= i\left(\bs{1} \partial_l + \bs{\Gamma}_l \right) \bs{\tPsi}\\
\bs{\Gamma}_l &:= -\frac{k_l}{2 \kn^2}  \bs{1} + \bs{A}_l\\
\bs{A}_l &:= \bs{E} \partial_l \bs{E}^{-1}\\
(\bs{E})_{j \mu} &:= E_{\mu j}
\end{split}
\end{equation}
The formulas (\ref{posop1}) or (\ref{posop2}) define the Hawton position operator for the photon when the Białynicki-Birula scalar product (\ref{bbsp}) is assumed to apply \cite{hawton,hawton2,hawton3}. Performing analogous calculations one can easily show that under the assumption that the scalar product has the form
\begin{equation}
\label{bbsp2}
\langle \bs{\tPhi} | \bs{\tPsi} \rangle' \sim  \int \frac{d^3 k}{(2 \pi)^3 \kn^{2s}} \bs{\tPhi}^{\hrm} \ofk \bs{\tPsi} \ofk,
\end{equation}
the formula (\ref{gcond}) has to read now
\begin{equation}
 \kn^{-s} g_{\mu \nu} = \mathrm{const}_{\mu \nu}
\end{equation}
and (\ref{Edef}) goes into
\begin{equation}
\label{Edef2}
\begin{split}
\vec{e}_{\mu} \ofk &= \kn^{s} \vec{E}_{\mu} \ofk \\
\vec{e}^{\, \mu} \ofk &= \kn^{-s} \vec{E}^{\,\mu} \ofk
\end{split}
\end{equation}
Consequently, the photon position operator in this case reads
\begin{equation}
\label{genposop}
\hat{X}_l \bs{\tPsi} = i\left(\bs{1} \partial_l -s \frac{k_l}{ \kn^2}  \bs{1} + \bs{A}_l \right) \bs{\tPsi}\\
\end{equation}
Thus we recover the general form of the photon position operator given by Margaret Hawton \cite{hawton}. One can quickly show that the action of $\hat{X}_l$ on $\bs{\tPsi}$ can be written in a compact form
\begin{equation}
\left(\hat{X}_l \bs{\tPsi} \right)_j = i e_{\mu j} \partial_l \left( \tensor{e}{^{\mu}_n} \tPsi_n \right)
\end{equation}
This corresponds to the last formula of section V in Hawton's work \cite{hawton}.

We are going now to study the form of connection $\dcon$ in more detail and to find an explicit expression for $\dcon$. First, we introduce some new coordinate system on $\rtt = \mathbb{R}^3 \setminus \{ (0,0,k_3) \in \mathbb{R}^3 : k_3 \geq 0\}$. Let $(k_1, k_2, k_3) \in \rtt$. We project this point on the unit $2$-sphere $S^2$ with center $(0,0,0)$ along the ray defined by $\vec{k}=(k_1,k_2,k_3)$. This projection determines the point $\vec{k}'$ on $S^2$ of the Cartesian coordinates $\left(\frac{k_1}{\kn},\frac{k_2}{\kn},\frac{k_3}{\kn}\right)$. Let $(\xi,\eta)$ be the stereographic coordinates of the point $\vec{k}'$ for the stereographic projection of $S^2$ from the north pole $(0,0,1)$ on the projection $2$-plane $k_3 = 0$. Finally, to the original point of the Cartesian coordinates $(k_1,k_2,k_3)$ one assigns the coordinates $(\xi,\eta,\zeta)$, where $\zeta:=\kn > 0$. Thus we construct the new coordinate system on $\rtt$. From the well known theory of stereographic projection one easily infers the relations between the Cartesian coordinates $(k_1,k_2,k_3)$ and the coordinates $(\xi,\eta,\zeta)$
\begin{subequations}
\begin{align}
\xi &= \frac{k_1}{\kn - k_3} & \eta &= \frac{k_2}{\kn - k_3} & \zeta &= \kn
\\
k_1 &= \zeta \frac{2 \xi}{\xi^2 + \eta^2 +1} & k_2 &= \zeta \frac{2 \eta}{\xi^2 + \eta^2 +1}& k_3 &= \zeta \frac{\xi^2 +\eta^2 -1}{\xi^2 + \eta^2 +1}
\end{align}
\end{subequations}
for $(\xi, \eta) \in \mathbb{R}^2$, $\zeta > 0$. The natural basis of the vector fields on $\rtt$ defined by the coordinates $(\xi,\eta,\zeta)$ is given by
\begin{subequations}
\label{natbas}
\begin{equation}
\begin{split}
\frac{\partial}{\partial \xi} &= \kn^{-1} \left[ \left( \kn (\kn -k_3)-k_1^2 \right) \frac{\partial}{\partial k_1} -k_1 k_2 \frac{\partial}{\partial k_2} + k_1 (\kn -k_3) \frac{\partial}{\partial k_3}\right]
\\ 
&= \frac{2 \zeta}{(\xi^2 + \eta^2 +1)^2} \left[ (\eta^2 - \xi^2+1) \frac{\partial}{\partial k_1} - 2\xi \eta \frac{\partial}{\partial k_2}  + 2\xi \frac{\partial}{\partial k_3}\right]
\\ 
&= 2 \kn \sin^2 \frac{\theta}{2}\Bigg[ \left(1-2 \cos^2 \frac{\theta}{2} \cos^2 \varphi  \right) \frac{\partial}{\partial k_1} 
-\cos^2 \frac{\theta}{2} \sin 2\varphi \frac{\partial}{\partial k_2} \\
& \quad + \sin \theta \cos \varphi \frac{\partial}{\partial k_3} \Bigg]
\end{split}
\end{equation}
\begin{equation}
\begin{split}
\frac{\partial}{\partial \eta} &= \kn^{-1} \left[-k_1 k_2 \frac{\partial}{\partial k_1} + \left( \kn (\kn -k_3)-k_2^2 \right) \frac{\partial}{\partial k_2} + k_2 (\kn -k_3) \frac{\partial}{\partial k_3}\right]
\\ 
&= \frac{2 \zeta}{(\xi^2 + \eta^2 +1)^2} \left[- 2\xi \eta \frac{\partial}{\partial k_1} + (\xi^2 - \eta^2+1) \frac{\partial}{\partial k_2}  + 2\eta \frac{\partial}{\partial k_3}\right]\\
&= 2 \kn \sin^2 \frac{\theta}{2}\Bigg[  -\cos^2 \frac{\theta}{2} \sin 2\varphi \frac{\partial}{\partial k_1} 
+\left(1-2 \cos^2 \frac{\theta}{2} \sin^2 \varphi  \right)\frac{\partial}{\partial k_2} \\
& \quad + \sin \theta \sin \varphi \frac{\partial}{\partial k_3} \Bigg]
\end{split}
\end{equation}
\begin{equation}
\begin{split}
\frac{\partial}{\partial \zeta} &= \kn^{-1} \left[k_1 \frac{\partial}{\partial k_1} + k_2 \frac{\partial}{\partial k_2} + k_3 \frac{\partial}{\partial k_3}\right]
\\ 
&= \frac{1}{(\xi^2 + \eta^2 +1)} \left[ 2\xi  \frac{\partial}{\partial k_1} + 2\eta \frac{\partial}{\partial k_2}  + (\xi^2+ \eta^2 -1) \frac{\partial}{\partial k_3}\right]\\
&=  \sin \theta \cos \varphi \frac{\partial}{\partial k_1} +  \sin \theta \sin \varphi \frac{\partial}{\partial k_2} + \cos \theta  \frac{\partial}{\partial k_3} 
\end{split}
\end{equation}
\end{subequations}
where $(\theta,\varphi)$, $0 < \theta \leq \pi$, $0 \leq \varphi < 2\pi$, are the spherical coordinates of the point $(k_1,k_2,k_3)$ or, equivalently, of the point $\left(\frac{k_1}{\kn},\frac{k_2}{\kn},\frac{k_3}{\kn}\right)$. (Note that for $\theta=\pi$ the coordinate $\varphi$ is undefined, but since $\sin \pi =0$ the formulas (\ref{natbas}) hold true also for $\theta =\pi$). One quickly finds that the vectors $\frac{\partial}{\partial \xi}$, $\frac{\partial}{\partial \eta}$ and $\frac{\partial}{\partial \zeta}$ are mutually orthogonal. Hence the coordinate system $(\xi, \eta, \zeta )$ is orthogonal. However, the most important advantage of this new system is the fact that
\begin{equation}
\frac{\partial}{\partial \xi}, \frac{\partial}{\partial \eta} \in \Pi^{\bot} \ofk
\end{equation}
for every $\vec{k} = (k_1,k_2,k_3) \in \rtt$. Consequently, one easily constructs an orthonormal triad of vector fields $(\vec{E}_1 \ofk, \vec{E}_2 \ofk, \vec{E}_3 \ofk)$ on $\rtt$ satisfying the conditions (\ref{orthonormalt}). It reads
\begin{equation}
\label{triad1}
\begin{split}
\vec{E}_1 \ofk &:= -(\kn -k_3)^{-1} \frac{\partial}{\partial \xi}= -\frac{\xi^2 + \eta^2 +1}{2\zeta} \frac{\partial}{\partial \xi}\\
\vec{E}_2 \ofk &:= (\kn -k_3)^{-1} \frac{\partial}{\partial \eta}= \frac{\xi^2 + \eta^2 +1}{2\zeta} \frac{\partial}{\partial \eta}\\
\vec{E}_3 \ofk &:= \frac{\partial}{\partial \zeta}
\end{split}
\end{equation}
(The system of coordinates $(\xi,\eta,\zeta)$ has the opposite orientation to the Cartesian system $(k_1,k_2,k_3)$. Hence the sign ``$-$'' in the formula defining $\vec{E}_1 \ofk$). Inserting (\ref{triad1}) into the definition of $\bs{A}_l$ given by (\ref{posop2}), after performing straightforward calculations we get
\begin{equation}
\label{Aform1}
i \bs{A}_l = \frac{(\vec{k} \times \vec{\bs{S}})_l}{\kn^2} + \epsilon_{lm3} \frac{k_m}{\kn (\kn -k_3)} \bs{\Sigma}
\end{equation}
where $\vec{\bs{S}}=(\bs{S}_1,\bs{S}_2,\bs{S}_3)$ and
\begin{align}
\label{sdef}
\bs{S}_1 &= 
\begin{pmatrix}
0 & 0 & 0\\
0 & 0 & -i\\
0 & i & 0 
\end{pmatrix},
&
\bs{S}_2 &= 
\begin{pmatrix}
0 & 0 & i\\
0 & 0 & 0\\
-i & 0 & 0 
\end{pmatrix},
&
\bs{S}_3 &= 
\begin{pmatrix}
0 & -i & 0\\
i & 0 & 0\\
0 & 0 & 0 
\end{pmatrix}
\end{align}
are the spin-$1$ matrices, and
\begin{equation}
\bs{\Sigma} = \frac{\vec{k} \cdot \vec{\bs{S}}}{\kn} =  i \kn^{-1}  
\begin{pmatrix}
0 & -k_3 & k_2\\
k_3 & 0 & -k_1\\
-k_2 & k_1 & 0 
\end{pmatrix}
\end{equation}
is \emph{the helicity operator}. Thus the photon position operator (\ref{posop2}) takes now the form 
\begin{equation}
\hat{X}_l = i\bs{1} \partial_l  -i \frac{k_l}{2 \kn^2}  \bs{1} + \frac{(\vec{k} \times \vec{\bs{S}})_l}{\kn^2} + \epsilon_{lm3} \frac{k_m}{\kn (\kn -k_3)} \bs{\Sigma}
\end{equation} 
and for general $s$ (see (\ref{genposop})) one has
\begin{equation}
\label{genposop2}
\hat{X}_l = i\bs{1} \partial_l  -i s \frac{k_l}{ \kn^2}  \bs{1} + \frac{(\vec{k} \times \vec{\bs{S}})_l}{\kn^2} + \epsilon_{lm3} \frac{k_m}{\kn (\kn -k_3)} \bs{\Sigma}
\end{equation} 
An analogous result to (\ref{genposop2}) has been found in \cite{hawton3} for the case when instead of $\mathbb{R}^3 \setminus \{ (0,0,k_3) \in \mathbb{R}^3 : k_3 \geq 0\}$ one takes $\mathbb{R}^3 \setminus \{ (0,0,k_3) \in \mathbb{R}^3 : k_3 \leq 0\}$. In \cite{hawton3} this result has been obtained from the photon position operator for the orthonormal triad defined by the spherical coordinates, with the use of the suitable rotation of this triad around the vector $\vec{k}$ (see the example in the present section). 

Now we are able to get the more general formula for the position operator of the photon. First, note that the general real orthonormal triad of vector fields on $\Omega \subset \rtt$ fulfilling the conditions (\ref{orthonormalt}) has the form
\begin{equation}
\label{realrot}
\begin{split}
\vec{E}^{\, '}_1 &= a \vec{E}_1 -b \vec{E}_2\\
\vec{E}^{\, '}_2 &= b \vec{E}_1 +a \vec{E}_2\\
\vec{E}^{\, '}_3 &= \vec{E}_3
\end{split}
\end{equation}
for $a=a \ofk$, $b=b \ofk$, $a^2 + b^2=1$ and $\vec{k} \in \Omega$. From (\ref{realrot}), employing (\ref{Aform1}), (\ref{triad1}) with (\ref{natbas}) making simple manipulations one finds
\begin{equation}
i \bs{A}'_l = 
i  \bs{E}' \partial_l {\bs{E}'}^{-1}
=\frac{(\vec{k} \times \vec{\bs{S}})_l}{\kn^2} + \left[ \epsilon_{lm3} \frac{k_m}{\kn (\kn -k_3)} +( a\partial_l b - b \partial_l a ) \right]\bs{\Sigma} 
\end{equation}
Hence, the more general form of the photon position operator reads
\begin{equation}
\label{posopgen}
\begin{split}
\hat{X}_l  = i\bs{1} \partial_l  -i s \frac{k_l}{ \kn^2}  \bs{1} +  \frac{(\vec{k} \times \vec{\bs{S}})_l}{\kn^2} + \left[ \epsilon_{lm3} \frac{k_m}{\kn (\kn -k_3)} +( a\partial_l b - b \partial_l a ) \right]\bs{\Sigma} \\
a=a \ofk, \quad b=b \ofk, \quad a^2 + b^2=1,  \quad \vec{k} \in \Omega \subset \rtt
\end{split}
\end{equation}
In the case of Białynicki-Birula  scalar product (\ref{bbsp}) one puts $s=\frac{1}{2}$. Note that the first three terms on the right side of (\ref{posopgen}) with $s=\frac{1}{2}$ define the position operator of the photon proposed by M.~H.~L.~Pryce in his pioneering work \cite{pryce}. However, the components of Pryce's position operator do not commute.

\begin{rmrk}
\label{unitrmrk}
To emphasize a geometric meaning of our construction we have decided to follow the path of real Riemannian geometry. For this reason the connection $\dcon$ is originally defined as the connection in the tangent bundle $T(\rtt)$ and then extended to its complexification, yielding in turn photon position operator $\hat{\vec{X}}$. It should be noted however, that in the general approach the connection coefficients $\Gamma_{ijk}$ may be complex valued. This can be taken into account by allowing unitary (instead of orthogonal) transformations of the triad $(\vec{E}_1,\vec{E}_2,\vec{E}_3)$ in (\ref{realrot}). To this end consider
\begin{align}
\bs{E}' &=\bs{E} \bs{U}\\
\bs{U} &= 
\begin{pmatrix}
\multicolumn{2}{r}{\bs{U}_{\bot}}& \begin{matrix}0\\0 \end{matrix}\\
0 &0 &1
\end{pmatrix}
\end{align}
with some unitary $2 \times 2$ matrix $\bs{U}_{\bot} \ofk$. After straightforward calculations one obtains
\begin{equation}
\label{aprgen}
 \bs{A}'_l =  \bs{E}' \partial_l {\bs{E}'}^{-1} =  \bs{A}_l +  \bs{E} \bs{U} \big( \partial_l \bs{U}^{\hrm} \big) \bs{E}^{T} 
=  \bs{A}_l +  \bs{E_{\bot}} \bs{U_{\bot}} \big( \partial_l \bs{U}_{\bot}^{\hrm} \big) \bs{E}_{\bot}^{T} 
\end{equation}
where $\bs{E}_{\bot}$ is $3 \times 2$ matrix determined by vectors $\vec{E}_1$ and $\vec{E}_2$
\begin{align}
(\bs{E}_{\bot})_{j \mu} &= E_{\mu j} & j&=1,2,3 \quad \mu=1,2
\end{align}
Writing the general $\vec{k}$-dependent unitary $2 \times 2$ matrix as
\begin{equation}
\bs{U}_{\bot} = e^{i \beta}
\begin{pmatrix}
e^{i \psi} & 0\\
0 & e^{-i \psi}
\end{pmatrix}
\begin{pmatrix}
\cos \alpha & \sin \alpha\\
- \sin \alpha & \cos \alpha
\end{pmatrix}
\begin{pmatrix}
e^{i \Delta} & 0\\
0 & e^{-i \Delta}
\end{pmatrix}
\end{equation}
with real functions $\beta = \beta \ofk$, $\alpha = \alpha \ofk$, 	$\psi = \psi \ofk$  and $\Delta = \Delta \ofk$, the general form of anti-Hermitian matrix $\bs{U_{\bot}}\partial_l \bs{U}_{\bot}^{\hrm}$ can be calculated
\begin{equation}
\label{udugen}
\bs{U_{\bot}}\partial_l \bs{U}_{\bot}^{\hrm} = 
i
\begin{pmatrix}
 -\cos (2 \alpha ) \partial_l \Delta-\partial_l \beta-\partial_l \psi & e^{2i \psi } \Big(\sin (2 \alpha ) \partial_l \Delta+i \partial_l \alpha \Big) \\
 e^{-2 i \psi } \Big(\sin (2 \alpha ) \partial_l \Delta-i \partial_l \alpha \Big) & \cos (2 \alpha ) \partial_l \Delta-\partial_l \beta+\partial_l \psi 
\end{pmatrix}
\end{equation}
The relation (\ref{udugen}) expanded in terms of Pauli matrices $\bsv{\sigma}_j$ and substituted into (\ref{aprgen}) produces the following formula for photon position operator
\begin{multline}
\label{posopgamcplx}
\hat{X}_l = i\bs{1} \partial_l  -i s \frac{k_l}{ \kn^2}  \bs{1} + \frac{(\vec{k} \times \vec{\bs{S}})_l}{\kn^2} 
+
\big(\partial_l \beta\big) \bs{R}_0
- \big(\cos(2 \psi) \sin (2 \alpha) \partial_l \Delta - \sin(2 \psi) \partial_l \alpha \big) \bs{R}_1 \\
+ \left( \epsilon_{lm3} \frac{k_m}{\kn (\kn -k_3)} +\sin(2 \psi) \sin (2 \alpha) \partial_l \Delta + \cos(2 \psi) \partial_l \alpha \right) \bs{R}_2 \\
+ \big(\cos(2 \alpha) \partial_l \Delta + \partial_l \psi \big) \bs{R}_3
\end{multline} 
where 
\begin{align}
\bs{R}_j &:= \bs{E_{\bot}} \bsv{\sigma}_j \bs{E}_{\bot}^{T}  & j&=0,1,2,3
\end{align}
(We use the fact that $\bs{R}_2=\bs{\Sigma}$). Clearly, the formula (\ref{posopgen}) can be obtained from (\ref{posopgamcplx}) by choosing $\beta  = \psi = \Delta \equiv 0$, and setting $a=\cos \alpha$ and $b=\sin \alpha$.
\end{rmrk}

As an example of the real transformation (\ref{realrot}) we investigate the case when the orthonormal triad $(\vec{E}^{\, '}_1, \vec{E}^{\, '}_2, \vec{E}^{\, '}_3)$ is determined in a natural way by the spherical system of coordinates  $(\theta,\varphi,\kn)$, $0 < \theta < \pi$, $0 \leq \varphi <2 \pi$
\begin{equation}
\label{triad2}
\begin{split}
\vec{E}^{\, '}_1 &= \cos \theta \cos \varphi \frac{\partial}{\partial k_1} + \cos \theta \sin \varphi \frac{\partial}{\partial k_2} - \sin \theta \frac{\partial}{\partial k_3} \\
\vec{E}^{\, '}_2 &= - \sin \varphi \frac{\partial}{\partial k_1} + \cos \varphi \frac{\partial}{\partial k_2}  \\
\vec{E}^{\, '}_3 &= \sin \theta \cos \varphi \frac{\partial}{\partial k_1} + \sin \theta \sin \varphi \frac{\partial}{\partial k_2} + \cos \theta \frac{\partial}{\partial k_3} 
\end{split}
\end{equation}
From (\ref{natbas}) and (\ref{triad1}) one gets
\begin{equation}
\label{triad1b}
\begin{split}
\vec{E}_1 &= \left(2 \cos^2 \frac{\theta}{2} \cos^2 \varphi -1 \right) \frac{\partial}{\partial k_1} + \cos^2 \frac{\theta}{2} \sin 2\varphi \frac{\partial}{\partial k_2} - \sin \theta \cos \varphi \frac{\partial}{\partial k_3} \\
\vec{E}_2 &= - \cos^2 \frac{\theta}{2} \sin 2\varphi \frac{\partial}{\partial k_1} + \left(1- 2 \cos^2 \frac{\theta}{2} \sin^2 \varphi \right) \frac{\partial}{\partial k_2} + \sin \theta \sin \varphi \frac{\partial}{\partial k_3} \\
\vec{E}_3 &= \sin \theta \cos \varphi \frac{\partial}{\partial k_1} + \sin \theta \sin \varphi \frac{\partial}{\partial k_2} + \cos \theta \frac{\partial}{\partial k_3} 
\end{split}
\end{equation}
Substituting (\ref{triad2}) and (\ref{triad1b}) into (\ref{realrot}) we quickly conclude that
\begin{align}
\label{abexmpl}
a&=\cos \varphi =\frac{k_1}{\sqrt{k_1^2 + k_2^2}}, & b&=\sin \varphi   =\frac{k_2}{\sqrt{k_1^2 + k_2^2}}
\end{align}
Then Eq.\ (\ref{posopgen}) gives now (compare with \cite{hawton3})
\begin{equation}
\hat{X}_l  = i\bs{1} \partial_l  -i s \frac{k_l}{ \kn^2}  \bs{1} +  \frac{(\vec{k} \times \vec{\bs{S}})_l}{\kn^2} + \left( \epsilon_{lm3} \frac{k_m}{\kn (\kn -k_3)} +\partial_l \varphi \right)\bs{\Sigma} 
\end{equation}
or after simple direct calculations one gets
\begin{equation}
\label{posophawt}
\hat{X}_l  = i\bs{1} \partial_l  -i s \frac{k_l}{ \kn^2}  \bs{1} +  \frac{(\vec{k} \times \vec{\bs{S}})_l}{\kn^2} +  \epsilon_{lm3} \frac{k_m k_3}{\kn (k_1^2 +k_2^2)}  \bs{\Sigma} 
\end{equation}
This is just the photon position operator found by M.~Hawton in \cite{hawton}. The domain $\Omega \subset \rtt$ where the right side of (\ref{posophawt}) is nonsingular is defined as $\Omega = \mathbb{R}^3 \setminus \{ (0,0,k_3) \in \mathbb{R}^3 : k_3 \in \mathbb{R}^1\}$. As the next example consider the following transformation
\begin{align}
\label{abexmpl2}
a&=\cos 2 \varphi =\frac{k_1^2 - k_2^2}{k_1^2 + k_2^2}, & b&=\sin 2 \varphi   =\frac{2 k_1 k_2}{k_1^2 + k_2^2}
\end{align}
defined on the same $\Omega = \mathbb{R}^3 \setminus \{ (0,0,k_3) \in \mathbb{R}^3 : k_3 \in \mathbb{R}^1\}$. Inserting this into (\ref{posopgen}) one quickly gets (see also \cite{hawton3})
\begin{equation}
\label{posopv3}
\hat{X}_l = i\bs{1} \partial_l  -i s \frac{k_l}{\kn^2}  \bs{1} + \frac{(\vec{k} \times \vec{\bs{S}})_l}{\kn^2} - \epsilon_{lm3} \frac{k_m}{\kn (\kn +k_3)} \bs{\Sigma}
\end{equation} 
Observe that $\hat{X}_l$ given by (\ref{posopv3}) is nonsingular on the open submanifold of $\mathbb{R}^3$ defined as $\mathbb{R}^3 \setminus \{ (0,0,k_3) \in \mathbb{R}^3 : k_3 \leq 0\}$. Straightforward calculations show that we can arrive at (\ref{posopv3}) by using the stereographic projection of the unit $2$-sphere from the south pole $(0,0,-1)$ and not from the north pole $(0,0,1)$ as it has been done in the case of the formula (\ref{genposop2}).

We can furthermore analyze a question of torsion of connection $\dcon$. The connection coefficients corresponding to the formula (\ref{posopgen}) read
\begin{equation}
\Gamma_{jml}= -\frac{1}{\kn^2} \left( s \delta_{jm} k_l - \delta_{ml} k_j + \delta_{jl} k_m + \epsilon_{rjm} \epsilon_{lp3} \frac{ k_p k_r}{\kn - k_3} + \kn \epsilon_{rjm} k_r \partial_l \alpha \right)
\end{equation}
for $a \ofk = \cos \alpha \ofk$ and $b \ofk = \sin \alpha \ofk$. From this expression it can be easily observed that no choice of $\alpha \ofk$ can make torsion $Q_{jml}=\Gamma_{jml}-\Gamma_{jlm}$ vanishing.  Indeed, the condition $Q_{112}=0$ gives
\begin{equation}
\label{tphi1}
\partial_1 \alpha = \frac{(s-1) k_2}{\kn k_3} - \frac{k_2}{k(\kn-k_3)}
\end{equation}
while $Q_{113}=0$ yields
\begin{equation}
\partial_1 \alpha = \frac{(1-s) k_3}{\kn k_2} - \frac{k_2}{k(\kn-k_3)}
\end{equation}
These equations are immediately inconsistent for $s \neq 1$. For $s=1$ it is enough to consider $Q_{221}=0$ producing
\begin{equation}
\label{tphi2}
\partial_2 \alpha = \frac{(1-s) k_1}{\kn k_3} + \frac{k_1}{k(\kn-k_3)}
\end{equation}
As it can be verified by direct calculation equations (\ref{tphi1}) and (\ref{tphi2}) do not satisfy basic integrability condition $\partial_1 \partial_2 \alpha = \partial_2 \partial_1 \alpha$. Thus, we conclude that the connection defining photon position operator (\ref{posopgen}) must have a non-vanishing torsion tensor. 

It can be interesting and informative to find the phase space image of the position operator $\hat{\vec{X}}$. To this end we use extensively the formalism developed in \cite{przanowski}. In line with that formalism photon phase space is given as
\begin{equation}
\left\{ \left( \vec{p}, \vec{x}, \phi_m, n \right) \right\} = \mathbb{R}^3 \times \mathbb{R}^3 \times \Gamma^3
\end{equation}
where $\Gamma^3$ is the $3 \times 3$ grid, $\Gamma^3 = \{ ( \phi_m, n ) \}$, with $m,n = 0,1,2$, $\phi_m = \frac{2 \pi}{3} m$. We assume that the kernels $\mathcal{P}\Big(\frac{\hbar \vec{\lambda} \cdot \vec{\mu} }{2} \Big)$, for $\vec{\lambda}, \vec{\mu} \in \mathbb{R}^3$ and $\mathcal{K} \Big( \frac{\pi k l}{3} \Big)$, for $k,l =0,1,2$, determining the \emph{Stratonovich-Weyl quantizer} (the \emph{Fano operators}) are taken as
\begin{equation}
\label{kerdef}
\mathcal{P}\left(\frac{\hbar \vec{\lambda} \cdot \vec{\mu} }{2} \right)=1, \quad \quad \mathcal{K} \left( \frac{\pi k l}{3} \right) = (-1)^{kl}
\end{equation}
(see Eqs.\ (5.8) and (6.7) of Ref.\ \cite{przanowski}). Then the phase space image of the position operator $\hat{\vec{X}}$ is given by the function (see Eq.\ (5.16) of Ref.\ \cite{przanowski})
\begin{equation}
\label{Xphsdef}
\vec{X} \big(\vec{p},  \vec{x}, \phi_m, n \big) = \Tr \left\{ \hat{\vec{X}} \hat{H}^{1/2} \hat{\Omega} \big(\vec{p},  \vec{x}, \phi_m, n \big) \hat{H}^{-1/2} \right\}
\end{equation}
where 
\begin{equation}
\hat{H}=c |\hat{\vec{p}}\,|=c \hbar |\hat{\vec{k}}|
\end{equation}
is the Hamiltonian operator, and $ \hat{\Omega} \big(\vec{p},  \vec{x}, \phi_m, n \big)$ is the \emph{Stratonovich-Weyl quantizer} for the kernels given by (\ref{kerdef}). This quantizer is defined as
\begin{multline}
\label{omegadef}
\hat{\Omega} \big(\vec{p},  \vec{x}, \phi_m, n \big) 
= \left(\frac{\hbar}{2 \pi}\right)^3 \frac{1}{3} \sum_{k,l=0}^2 \int d^3 \lambda d^3 \mu \, (-1)^{kl} \exp \left\{-i \left( \vec{\lambda} \cdot \vec{p} + \vec{\mu} \cdot \vec{x} \right) \right\} 
\times\\
\times \exp \left\{ -i \left( k \phi_m + \phi_l n\right)\right\} \hat{\mathcal{U}} (\vec{\lambda},\vec{\mu}) \otimes \hat{\mathcal{D}} (k,l)
\end{multline}
with
\begin{subequations}
\label{UDdef}
\begin{equation}
\begin{split}
\hat{\mathcal{U}} (\vec{\lambda},\vec{\mu}) =  \exp \left\{i \left( \vec{\lambda} \cdot \hat{\vec{p}} + \vec{\mu} \cdot \hat{\vec{x}} \right) \right\} &= \int d^3 x \exp \left\{ i \vec{\mu} \cdot \vec{x} \right\} \Ket{\vec{x}-\frac{\hbar \vec{\lambda}}{2}} \Bra{\vec{x}+\frac{\hbar \vec{\lambda}}{2}} 
\\
&=\int d^3 p \exp \left\{ i \vec{\lambda} \cdot \vec{p} \right\} \Ket{\vec{p}+\frac{\hbar \vec{\mu}}{2}} \Bra{\vec{p}-\frac{\hbar \vec{\mu}}{2}}
\end{split}
\end{equation}
\begin{equation}
\begin{split}
\hat{\mathcal{D}} (k,l) &= \exp \left\{ -i \frac{\pi k l}{3} \right\} \exp \left\{ i k \hat{\phi} \right\} \exp \left\{ i \frac{2 \pi}{3} l \hat{n} \right\}\\
&=
\exp \left\{ i \frac{\pi k l}{3} \right\} \exp \left\{ i \frac{2 \pi}{3} l \hat{n} \right\} \exp \left\{ i k \hat{\phi} \right\} \\
&=\exp \left\{ i \frac{\pi k l}{3} \right\} \sum_{m=0}^2 \exp \left\{ i \frac{2 \pi k m}{3} \right\} \Ket{\phi_{(m+l) \bmod 3}} \Bra{\phi_m}\\
&=\exp \left\{ i \frac{\pi k l}{3} \right\} \sum_{n=0}^2 \exp \left\{ i \frac{2 \pi n l}{3} \right\} \Ket{n} \Bra{(n+k) \bmod 3},
\end{split}
\end{equation}
\begin{align*}
\hat{n} &= \sum_{n=0}^2 n \Ket{n} \Bra{n}, & \hat{\phi}&=\sum_{m=0}^2 \phi_m \Ket{\phi_m} \Bra{\phi_m}, & \Ket{\phi_m} &= \frac{1}{\sqrt{3}} \sum_{n=0}^2 \exp \left\{ i n \phi_m \right\} \Ket{n} 
\end{align*}
\end{subequations}
(see Eqs.\ (5.3), (5.4) and (5.8) of \cite{przanowski}; see also \cite{przanowski2}). We choose the basis $\left\{ \Ket{n} \right\}_{n=0,1,2}$ so that the representation of the spin-$1$ operator $\hat{\vec{\mathcal{S}}}$ with respect to this basis is given by (\ref{sdef}), i.e. $(\mathcal{S}_j)_{rl} = - i \epsilon_{jrl}$, $j,r,l = 1,2,3$. Inserting $\hat{\Omega} \big(\vec{p},  \vec{x}, \phi_m, n \big)$ given by (\ref{omegadef}) with (\ref{UDdef}) and $\hat{\vec{X}}$ defined by (\ref{posopgen}) with $s=\frac{1}{2}$ into (\ref{Xphsdef}), and performing straightforward but tedious manipulations one finds the components of vector function $\vec{X} \big(\vec{p},  \vec{x}, \phi_m, n \big)$ as
\begin{multline}
X_l \big(\vec{p},  \vec{x}, \phi_m, n \big) = \\
x_l + 2\hbar \Bigg[ \frac{p_{((n+2) \bmod 3) +1}}{|\vec{p}\,|^2} \delta_{l, ((n+1) \bmod 3) +1} 
- \frac{p_{((n+1) \bmod 3) +1}}{|\vec{p}\,|^2} \delta_{l, ((n+2) \bmod 3) +1} \\
+ \Bigg( \frac{p_j}{|\vec{p}\,|(|\vec{p}\,|-p_3)}\epsilon_{lj3}+a \frac{\partial b}{\partial p_l} - b \frac{\partial a}{\partial p_l} \Bigg) \frac{p_r}{|\vec{p}\,|} \epsilon_{r, ((n+1) \bmod 3) +1, ((n+2) \bmod 3) +1} \Bigg] \sin \phi_m
\end{multline}
for $l=1,2,3$; $m,n=0,1,2$, $\phi_m = \frac{2 \pi}{3} m$ and with summation over $j,r=1,2,3$. Observe that the phase space image $\vec{X} \big(\vec{p},  \vec{x}, \phi_m, n \big)$ of the position operator $\hat{\vec{X}}$ depends not only on $\vec{x}$ but also on $\vec{p}$ and on the grid $\Gamma^3$ coordinates $(\phi_m,n)$. Namely, it is equal to $\vec{x}$ plus a term linear in $\hbar$ dependent on $(\vec{p}, \phi_m, n)$.

\section{Eigenfunctions of $\hat{\vec{X}}$}

Now, we are going to show how the geometrical interpretation of the Hawton position operator for the photon enables us to find in an easy way the respective eigenfunctions. These eigenfunctions have been introduced in \cite{hawton2,hawton3,debierre,hawton6}. 

Employing our results one infers from (\ref{dezero}) and (\ref{Edef}) that
\begin{align}
\hat{D}_l \left( \kn^{\frac{1}{2}} \vec{E}_{\mu} \right)&=0 & \mu&=1,2,3
\end{align}
Hence 
\begin{equation}
\hat{D}_l \left( \exp\left(- i \vec{k} \cdot \vec{X} \right)  \kn^{\frac{1}{2}} \vec{E}_{\mu} \right) = -i X_l \exp\left(- i \vec{k} \cdot \vec{X} \right)  \kn^{\frac{1}{2}} \vec{E}_{\mu} 
\end{equation}
Finally, as $\hat{X}_l = i \hat{D}_l$ we get
\begin{equation}
\hat{X}_l \left( \exp\left(- i \vec{k} \cdot \vec{X} \right)  \kn^{\frac{1}{2}} \vec{E}_{\mu} \right) = X_l \exp\left(- i \vec{k} \cdot \vec{X} \right)  \kn^{\frac{1}{2}} \vec{E}_{\mu} 
\end{equation}
for $\mu =1,2,3$. Since for any photon state the condition (\ref{ortcond}) must be fulfilled, the position eigenfunction of the photon $\bs{\tPsi}_{\vec{X}} \ofk$ has the form
\begin{equation}
\label{eigenf}
\bs{\tPsi}_{\vec{X}} \ofk = \left(c_1 \bs{E}_1 \ofk + c_2 \bs{E}_2 \ofk\right) \exp\left(- i \vec{k} \cdot \vec{X} \right)  \kn^{\frac{1}{2}}
\end{equation}
where $c_1,c_2 \in \mathbb{C}^1$, while $\bs{E}_1 \ofk$ and $\bs{E}_2 \ofk$ are the one-column matrices representing the vectors $\vec{E}_1 \ofk$ and $\vec{E}_2 \ofk$, respectively. Then the Białynicki-Birula scalar product (\ref{bbsp}) of the position eigenfunctions is normalized to the Dirac delta iff $|c_1|^2 + |c_2|^2 =1$
\begin{equation}
|c_1|^2 + |c_2|^2 =1 \Longleftrightarrow \int \volel \bs{\tPsi}^{\hrm}_{\vec{X}} \ofk \bs{\tPsi}_{\vec{X}'} \ofk = \delta (\vec{X} - \vec{X}')
\end{equation}
In particular taking $c_1 = \frac{1}{\sqrt{2}}$, $c_2 = \pm \frac{i}{\sqrt{2}}$ one has
\begin{equation*}
\bs{\tPsi}_{\vec{X}, \pm 1}\ofk  = \frac{1}{\sqrt{2}} \left( \bs{E}_1 \ofk \pm i \bs{E}_2 \ofk \right)  \exp\left(- i \vec{k} \cdot \vec{X} \right)  \kn^{\frac{1}{2}}
\end{equation*}
\begin{align}
\hat{\vec{X}} \bs{\tPsi}_{\vec{X}, \pm 1} &= \vec{X} \bs{\tPsi}_{\vec{X}, \pm 1} & \bs \Sigma \bs{\tPsi}_{\vec{X}, \pm 1} &=  \pm \vec{X} \bs{\tPsi}_{\vec{X}, \pm 1}
\end{align}
Analogously when the scalar product (\ref{bbsp2}) applies we employ (\ref{Edef2}) and, consequently
\begin{equation}
\label{eigenf2}
\bs{\tPsi}_{\vec{X}} \ofk = \left(c_1 \bs{E}_1 \ofk + c_2 \bs{E}_2 \ofk\right) \exp\left(- i \vec{k} \cdot \vec{X} \right)  \kn^s
\end{equation}
for $c_1,c_2 \in \mathbb{C}^1$ (see \cite{hawton2,hawton3,debierre,hawton6}). 

Given $\bs{\tPsi}_{\vec{X}} \ofk$ by (\ref{eigenf}), using the results of \cite{bialynicki2,bialynicki3,przanowski} one finds the eigenfunction in $\vec{x}$-representation  $\bs{\Psi}_{\vec{X}} (\vec{x})$ as the Fourier transform of $\bs{\tPsi}_{\vec{X}} \ofk$
\begin{multline}
\label{xrep}
\bs{\Psi}_{\vec{X}} (\vec{x}) = \sqrt{\hbar c} \int \frac{d^3 k}{(2 \pi )^3} \bs{\tPsi}_{\vec{X}} \ofk \exp \left( i \vec{k} \cdot \vec{x}\right)\\ = 
\sqrt{\hbar c} \int \frac{d^3 k}{(2 \pi )^3} \left(c_1 \bs{E}_1 \ofk + c_2 \bs{E}_2 \ofk\right) \kn^{\frac{1}{2}} \exp\left( i \vec{k} \cdot (\vec{x}-\vec{X}) \right)  
\end{multline}
In the general case when  $\bs{\tPsi}_{\vec{X}} \ofk$ is given by (\ref{eigenf2}) one gets
\begin{equation}
\bs{\Psi}_{\vec{X}} (\vec{x}) \sim
\sqrt{\hbar c} \int \frac{d^3 k}{(2 \pi )^3} \left(c_1 \bs{E}_1 \ofk + c_2 \bs{E}_2 \ofk\right) \kn^s \exp\left( i \vec{k} \cdot (\vec{x}-\vec{X}) \right)  
\end{equation}
%
%
%
\begin{exmpl}
Assume that $\bs{E}_1 \ofk$ and $\bs{E}_2 \ofk$ are given by (\ref{triad1b}). Then (\ref{xrep}) in the matrix form reads
\begin{equation}
\label{JT-psi}
\begin{split}
\bs{\Psi}_{\vec{X}} (\vec{x})=&
\frac{ \sqrt{\hbar c}}{(2 \pi)^3}  \int_{0}^{\infty}dk \,{k}^{5/2}\int_{0}^{\pi}d \theta \sin \theta \int_{0}^{2 \pi}d \varphi\\
&\times
\renewcommand*{\arraystretch}{1.4}
\begin{pmatrix}
c_1 \big(2 \cos^2 \frac{\theta}{2} \cos^2 \varphi -1\big)+ c_2\big(-\cos^2 \frac{\theta}{2}  \sin 2 \varphi \big) \\
c_1 \big(\cos^2 \frac{\theta}{2}  \sin 2 \varphi\big) + c_2 \big(1- 2 \cos^2 \frac{\theta}{2}  \sin^2 \varphi \big)
\\
c_1 \big(- \sin \theta \cos \varphi \big) + c_2 \big( \sin \theta \sin \varphi\big)
\end{pmatrix}
\\
&\times \exp \Big(i k \sin \theta \cos \varphi (x_1 - X_1)  + i k \sin \theta \sin \varphi (x_2 - X_2) +  i k \cos \theta  (x_3 - X_3)\Big)
\end{split}
\end{equation}
One quickly recognizes that the integrand in (\ref{JT-psi}) goes to infinity for $k \to \infty$. To avoid this problem we proceed in a standard way. Namely, we multiply the integrand by $\exp(-\varepsilon k)$, $\varepsilon >0$, and after performing integration we take the limit $\varepsilon \to 0^+$. Note that the limit should be calculated in the sense of distribution theory. To do it all one can apply the Wolfram Mathematica. Then, without going into details (which will be considered in a separate work) we arrive at the following results. The wave function $\bs{\Psi}_{\vec{X}} (\vec{x})$ can be written as
\begin{equation}
\label{JT-psi2}
\bs{\Psi}_{\vec{X}} (\vec{x})=
\begin{pmatrix}
c_1 \mathcal{F}_{I} + c_2 \mathcal{F}_{II} \\
-c_1 \mathcal{F}_{II} + c_2 \big(\mathcal{F}_{I} + \mathcal{F}_{III} \big) \\
c_1 \mathcal{F}_{IV} + c_2 \mathcal{F}_{V} 
\end{pmatrix}
\end{equation}
where $\mathcal{F}_{I} = \mathcal{F}_{I}(\vec{x}), \dots, \mathcal{F}_{V} = \mathcal{F}_{V}(\vec{x})$ are distributions which (in the sense of distribution theory) are equal to the following functions on respective domains: 
\begin{equation}
\label{JT-F}
\begin{split}
\mathcal{F}_{I} = & \frac{ \sqrt{\hbar c}}{8 \sqrt{2} \pi^{3/2}X^{7/2}} \Bigg(\frac{2i \cos 2 \varphi_1 \cos \theta_1}{\sin^2 \theta_1} - \frac{2 \cos 2 \varphi_1}{\sin^2 \theta_1} + \frac{2 \cos 2 \varphi_1(1-i  {\rm sgn} (\cos \theta_1))}{|\cos \theta_1|^{3/2}\sin^2 \theta_1}\\
 &+ 3 \sin^2 \varphi_1 - 5i \cos \theta_1 \sin^2 \varphi_1 \Bigg), \quad \text{for } X \neq 0,\; \theta_1 \neq \frac{\pi}{2}\\
\mathcal{F}_{II} = &\frac{\sqrt{\hbar c} \sin 2 \varphi_1}{4 \sqrt{2}\pi^{3/2}X^{7/2}\sin^2 \theta_1}
 \Bigg(
 \frac{11}{8} - \frac{21 \cos \theta_1}{16}  - \frac{3 \cos 2 \theta_1}{8}  + \frac{5i \cos 3\theta_1}{16} \\
 & - \frac{1 - i \,{\rm sgn}(\cos \theta_1)}{|\cos \theta_1|^{3/2}} \Bigg), \quad \text{for } X \neq 0,\; \theta_1 \neq \frac{\pi}{2}\\
\mathcal{F}_{III} = & \frac{\sqrt{\hbar c}(-3-5i \cos \theta_1)}{8 \sqrt{2}\pi^{3/2} X^{7/2}}, \quad \text{for } X \neq 0\\
\mathcal{F}_{IV} = & \frac{-5i \sqrt{\hbar c} \sin \theta_1 \cos \varphi_1}{8 \sqrt{2} \pi^{3/2} X^{7/2}}, \quad \text{for } X \neq 0\\
\mathcal{F}_{V} = & \frac{5i \sqrt{\hbar c} \sin \theta_1 \sin \varphi_1}{8 \sqrt{2} \pi^{3/2} X^{7/2}},  \quad \text{for } X \neq 0
\end{split}
\end{equation}
where we use the abbreviation $X:= |\vec{x} - \vec{X}|$ and $\theta_1$, $\varphi_1$ are the angles defining the direction of the vector $\vec{x} - \vec{X}$
\begin{equation}
\frac{\vec{x} - \vec{X}}{|\vec{x} - \vec{X}|}= \Big( \sin \theta_1 \cos\varphi_1 , \sin \theta_1 \sin \varphi_1 , \cos \theta_1 \Big)
\end{equation}
It is evident that the wave function $\bs{\Psi}_{\vec{X}} (\vec{x})$, $\vec{x} \in \mathbb{R}^3$, is a one column matrix with elements being distributions given by regularization of the functions (\ref{JT-F}) \cite{gelfand,vilenkin}. We have not been able yet to find these distributions in a clear compact form. However, from the partial result given by Eq.\ (\ref{JT-F}) one can draw some interesting conclusion. From the Białynicki-Birula \cite{bialynicki2,bialynicki3} and Sipe \cite{sipe} interpretation of the photon wave function $\bs{\Psi}_{\vec{X}} (\vec{x})$, the quantity $\bs{\Psi}_{\vec{X}}^{\hrm} (\vec{x}) \bs{\Psi}_{\vec{X}} (\vec{x}) d^3 x$ (if it exists) is proportional to the probability that the energy of the photon is localized in the domain $d^3 x$. Using this interpretation one can state that formula (\ref{JT-psi2}) under (\ref{JT-F}) shows that the energy of photon in the quantum state $\bs{\Psi}_{\vec{X}} (\vec{x})$ is localized in a small region $|\vec{x} - \vec{X}| \to 0$, $\theta_1 \to \frac{\pi}{2}$. To illustrate this behavior of $\bs{\Psi}_{\vec{X}} (\vec{x})$ we present the figures (\ref{first-fig}) to (\ref{last-fig}).
\begin{figure}
\centering
\begin{subfigure}{.45\textwidth}
  \centering
  \includegraphics[width=.9\linewidth]{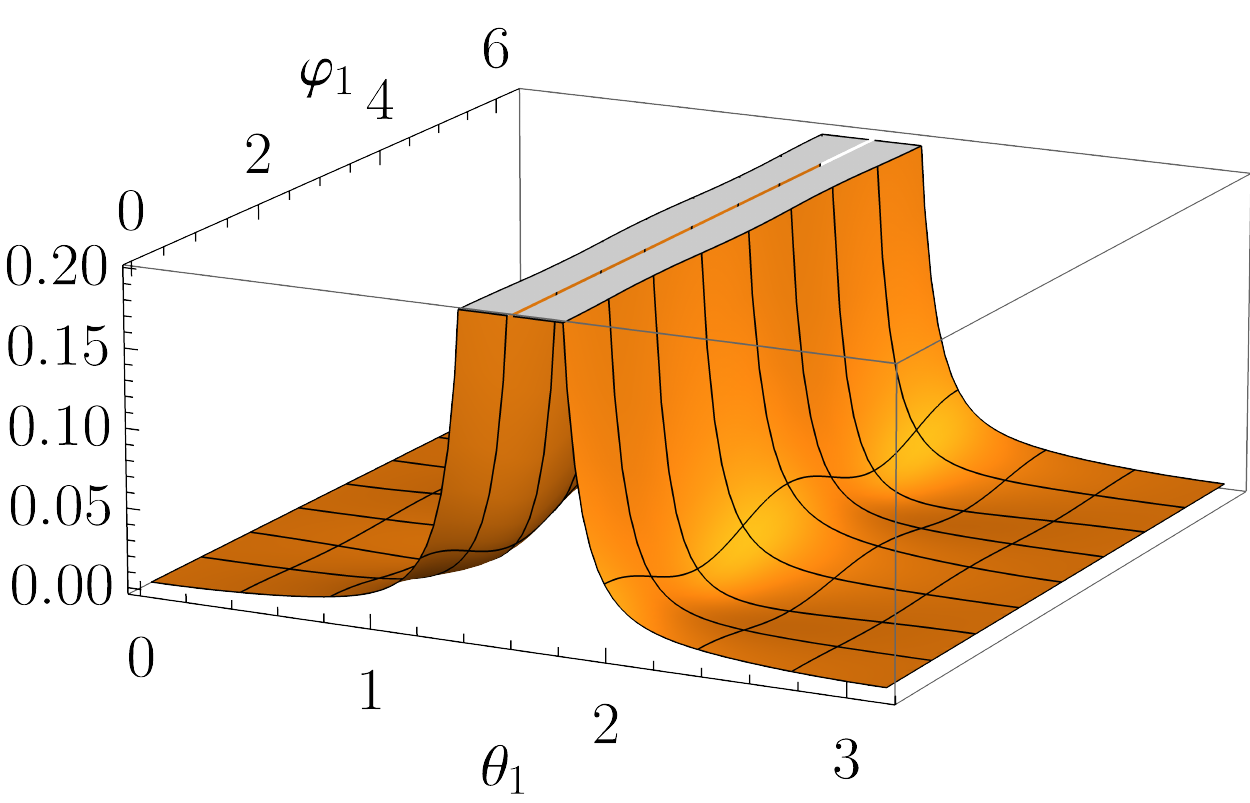}  
  \caption{$c_1=1$, $c_2=0$, $X=1$}
  \label{first-fig}
\end{subfigure}
\begin{subfigure}{.45\textwidth}
  \centering
  \includegraphics[width=.9\linewidth]{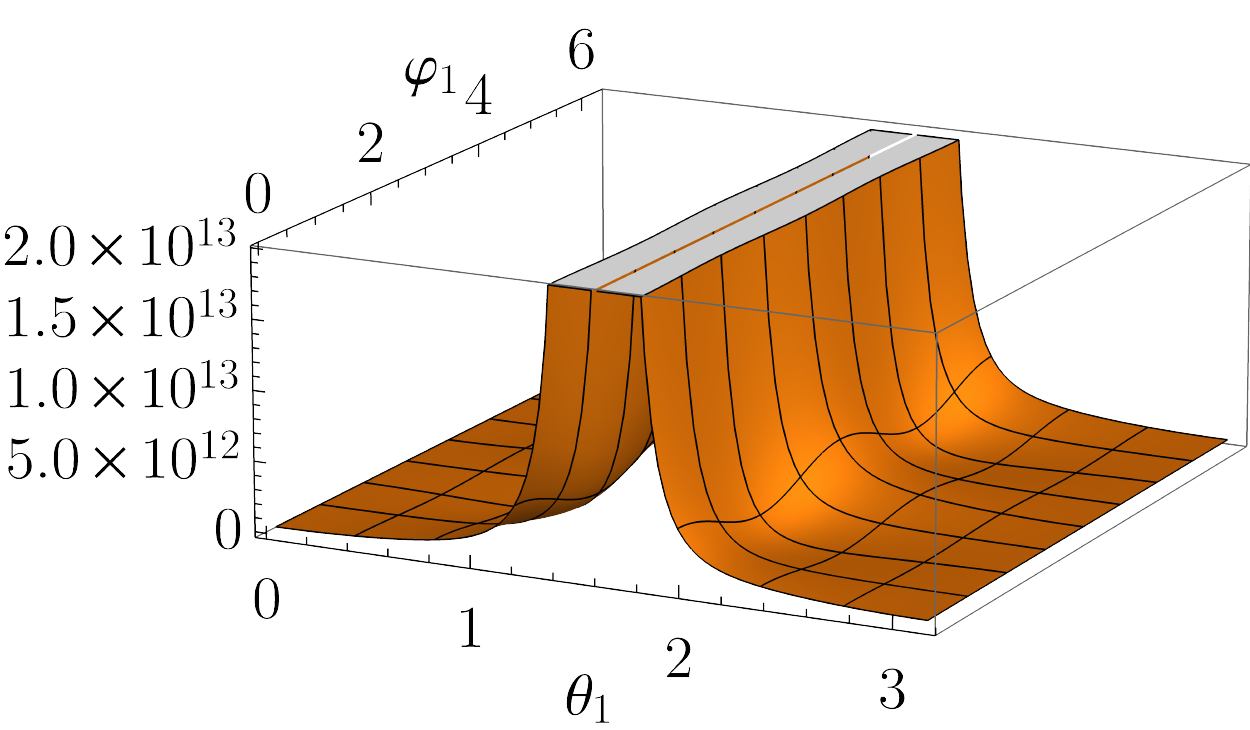}  
  \caption{$c_1=1$, $c_2=0$, $X=0.01$}
\end{subfigure}
\begin{subfigure}{.45\textwidth}
  \centering
  \includegraphics[width=.9\linewidth]{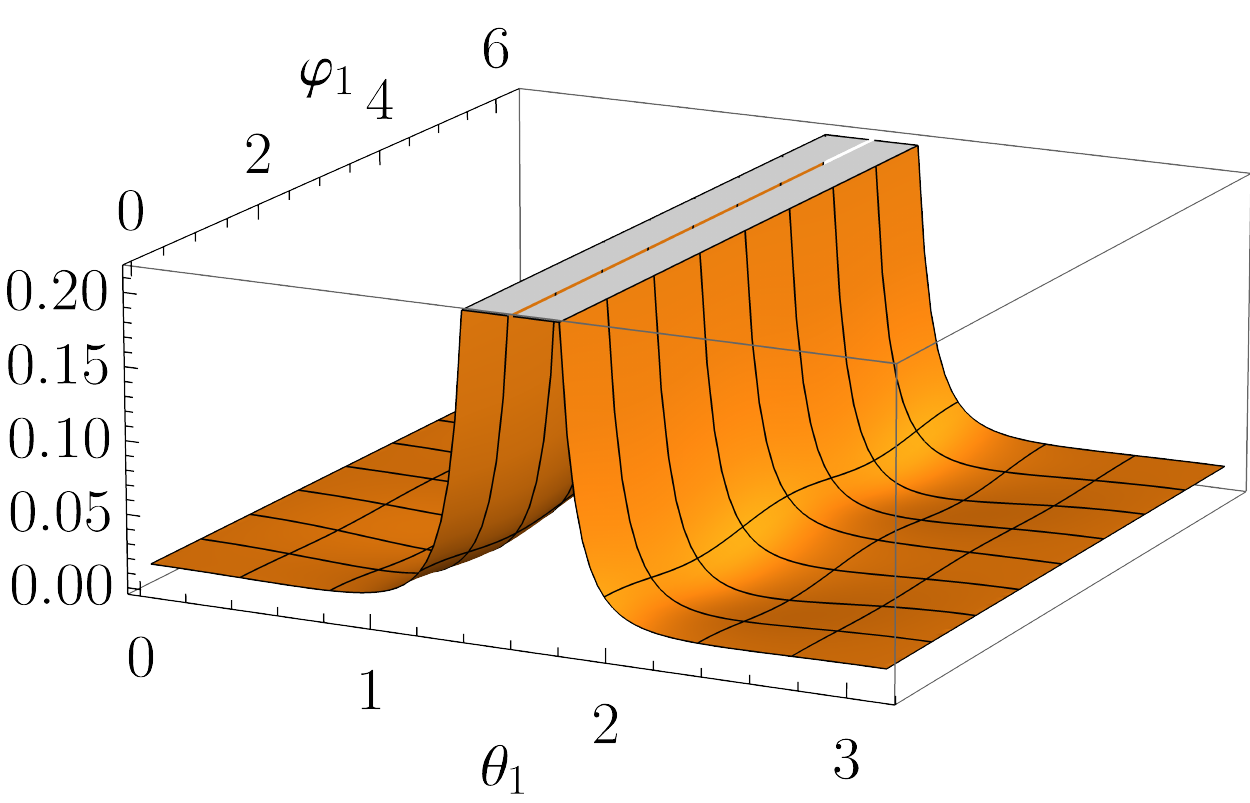}  
  \caption{$c_1=0$, $c_2=1$, $X=1$}
\end{subfigure}
\begin{subfigure}{.45\textwidth}
  \centering
  \includegraphics[width=.9\linewidth]{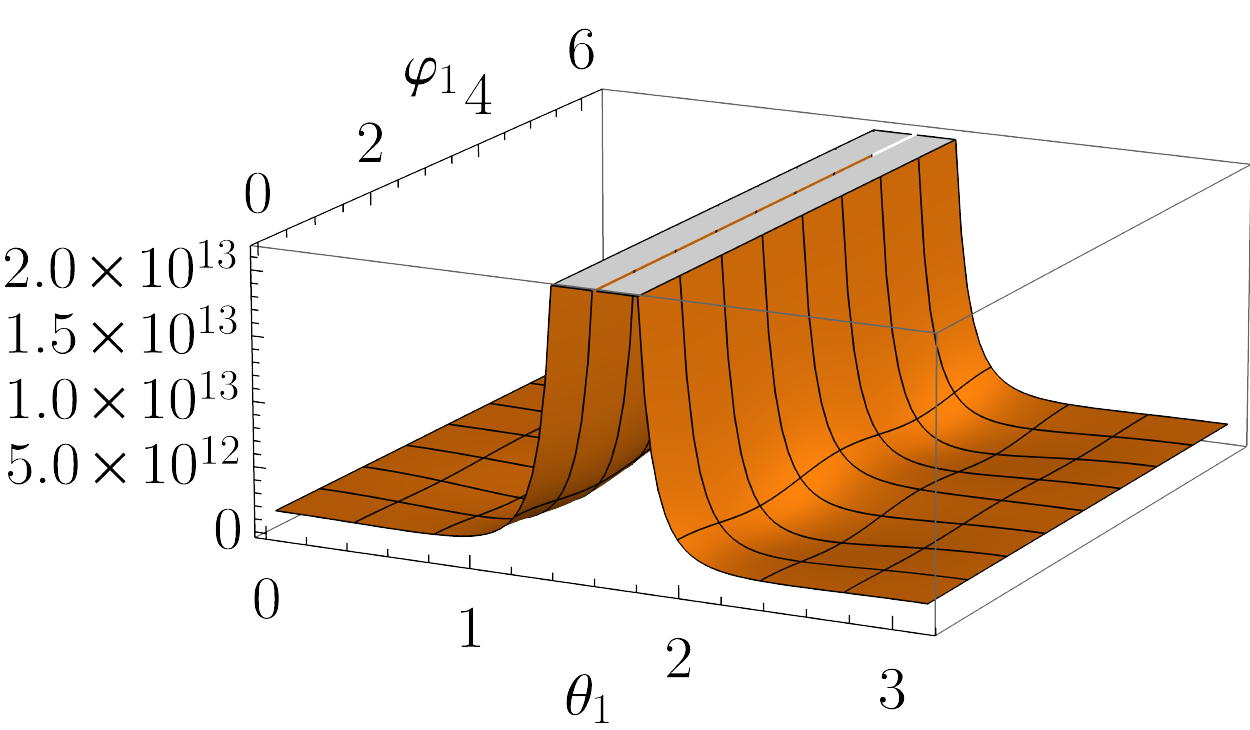}  
  \caption{$c_1=0$, $c_2=1$, $X=0.01$}
  \label{last-fig}
\end{subfigure}
\caption{Distribution of probability for photon energy, $\bs{\Psi}_{\vec{X}}^{\hrm} (\vec{x}) \bs{\Psi}_{\vec{X}} (\vec{x})$, as a function of $\theta_1$ and $\varphi_1$ for specific values of $X$ and $c_i$. For all cases $\hbar=1$ and $c=1$.}
\end{figure}

\end{exmpl}
%
%
%
Finally, it could be noted that this interpretation of $\bs{\Psi}_{\vec{X}} (\vec{x})$ is rather non-standard. The Fourier transform (\ref{xrep}) corresponds to Białynicki-Birula scalar product (\ref{bbsp}) of $\bs{\tPsi}_{\vec{X}} \ofk $ with functions
\begin{align}
\bs{\tPhi}_{1,\vec{x}} \ofk &\sim 
\begin{pmatrix}
\kn e^{-i \vec{k} \cdot \vec{x}}\\
0\\
0
\end{pmatrix} 
&
\bs{\tPhi}_{2,\vec{x}} \ofk &\sim 
\begin{pmatrix}
0\\
\kn e^{-i \vec{k} \cdot \vec{x}}\\
0
\end{pmatrix} 
&
\bs{\tPhi}_{3,\vec{x}} \ofk &\sim 
\begin{pmatrix}
0\\
0\\
\kn e^{-i \vec{k} \cdot \vec{x}}
\end{pmatrix} 
\end{align}
However these functions do not satisfy the condition (\ref{ortcond}), and they are not orthogonal in the sense of Białynicki-Birula scalar product. 
In turn, the ``orthodox'' quantum mechanical requirement of relating observables to self-adjoint operators has been abandoned here.  

\section{From the photon position operator to Berry's potential}

It has been noted by M. Hawton \cite{hawton} and then investigated further in \cite{hawton3} that the last term on the right side of Eq.~(\ref{posophawt}) defines some Berry potential leading to a Berry phase predicted by R.~Y.~Chiao and Y.~S.~Wu \cite{chiao} and confirmed experimentally by A.~Tomita and R.~Y.~Chiao \cite{tomita}. A deep group theoretical and geometrical interpretation of this Berry potential was given by I.~Białynicki-Birula and Z.~Białynicka-Birula \cite{bialynicki}. Here we will briefly repeat the problem by employing the general formula (\ref{posopgen}). 

Let us define a new covariant derivative (a connection) $\dcon' = (\dcp_1,\dcp_2,\dcp_3)$ on some domain $\Omega \subset \rtt$ \cite{bialynicki4,staruszkiewicz,bialynicki,bialynicki3,hawton,hawton3,debierre2}
\begin{equation}
\label{connprime}
\begin{split}
\dcp_l  := \partial_l  +i \left[ \epsilon_{lm3} \frac{k_m}{\kn (\kn -k_3)} +( a\partial_l b - b \partial_l a ) \right]\bs{\Sigma} \\
a=a \ofk, \quad b=b \ofk, \quad a^2 + b^2=1,  \quad \vec{k} \in \Omega \subset \rtt
\end{split}
\end{equation}
The curvature of $\dcon'$ is given as 
\begin{multline}
\left[ \dcp_l,\dcp_m \right]= i \epsilon_{lmr} \frac{k_r}{\kn^3} \bs{\Sigma} + i \left( \epsilon_{mr3} \frac{k_r}{\kn (\kn -k_3)} + a\partial_m b - b \partial_m a  \right) \frac{\left(\delta_{lj} -\frac{k_l k_j}{\kn^2}\right)}{\kn} \bs{S}_j \\
- i \left( \epsilon_{lr3} \frac{k_r}{\kn (\kn -k_3)} + a\partial_l b - b \partial_l a  \right) \frac{\left(\delta_{mj} -\frac{k_m k_j}{\kn^2}\right)}{\kn} \bs{S}_j
\end{multline}
To extract a Berry potential from the connection (\ref{connprime}) consider the photon with the helicity $\lambda = \pm 1$ moving so that the photon wave function $\bs{\tPsi} \ofk$ is parallelly propagated with respect to $\dcon'$. The momentum of photon changes along the curve 
\begin{align}
C: \vec{k} &= \vec{k} (\tau ) = (k_1 (\tau ), k_2 (\tau ),k_3 (\tau )), & \tau_0 &\leq \tau \leq \tau_1
\end{align}
Hence
\begin{equation}
\frac{d k_l}{d \tau} \dcp_l \bs{\tPsi} \big(\vec{k} (\tau)\big) = 0 \Longrightarrow \frac{d \bs{\tPsi}}{d \tau} +i \left(  \epsilon_{lm3} \frac{k_m \frac{d k_l}{d \tau}}{\kn (\kn -k_3)} + a \frac{d b}{d \tau} - b \frac{d a}{d \tau}  \right)  \bs{\Sigma} \bs{\tPsi}=0
\end{equation}
Then, since
\begin{equation}
\bs{\Sigma} \bs{\tPsi} = \lambda \bs{\tPsi}, \quad \quad  \lambda = \pm 1
\end{equation}
one arrives at the following ODE
\begin{equation}
\label{bode}
\frac{d \bs{\tPsi}}{d \tau} + i \lambda \left(\frac{k_2 \frac{d k_1}{d \tau}- k_1 \frac{d k_2}{d \tau}}{\kn (\kn -k_3)} + a \frac{d b}{d \tau} - b \frac{d a}{d \tau}\right) \bs{\tPsi}=0
\end{equation}
The solution of Eq.~(\ref{bode}) reads
\begin{align}
\label{bsol}
\bs{\tPsi} \big( \vec{k} (\tau) \big) &= \exp \left( i \int_C \mathcal{A}_l \,  dk_l\right) \bs{\tPsi}_0 \ofk & \bs{\tPsi}_0 \ofk &:= \bs{\tPsi} \big(\vec{k} (\tau_0) \big)
\end{align}
where
\begin{align}
\label{adef}
\vec{\mathcal{A}}&=(\mathcal{A}_1,\mathcal{A}_2,\mathcal{A}_3), & \mathcal{A}_l &:= -\lambda \left( \epsilon_{lm3} \frac{k_m}{\kn (\kn -k_3)} + a \partial_l b - b \partial_l a  \right)
\end{align}
is \emph{the Berry potential}.

If $C$ is a closed loop so that $\vec{k} (\tau_1)=\vec{k} (\tau_0) = \vec{k}$ one has from (\ref{bsol})
\begin{equation}
\label{bphase}
\begin{split}
\bs{\tPsi} \ofk & = \exp\Big( i \gmc{} \Big) \bs{\tPsi}_0 \ofk\\
\gmc{}& := \oint_C \mathcal{A}_l \,  dk_l
\end{split}
\end{equation}
and $\gmc{}$ is \emph{the Berry phase}.

We consider now some examples.
\begin{exmpl}
Assume that $\Omega = \rtt$ and $a= a \ofk$, $b = b \ofk$ are arbitrary differentiable functions on $\rtt$ satisfying the condition $a^2+b^2=1$ (see (\ref{posopgen})). From (\ref{adef}) and (\ref{bphase}) one gets
\begin{equation}
\gmc{1} = \lambda \oint_C \left( \frac{k_1 d k_2 - k_2 d k_1}{\kn (\kn -k_3)} + a\, db - b\, da \right)
\end{equation}
Then, since
\begin{equation}
a^2+b^2=1 \Longrightarrow a\, da + b\, db=0 \Longrightarrow da \wedge db =0
\end{equation}
we quickly obtain
\begin{equation}
d(a\, db - b\, db) = 2 da \wedge db=0
\end{equation}
Consequently, as the domain $\rtt$ is simply connected the Stokes theorem gives
\begin{equation}
\oint_C a\, db - b\, da = \int_S d(a\, db - b\, da)=0
\end{equation}
where $S$ is $2$-surface such that the loop $C$ is the boundary of $S$, $C=\partial S$. Finally $\gmc{1}$ is independent of $a$ and $b$, and it reads
\begin{equation}
\gmc{1} = \lambda \oint_C  \frac{k_1 d k_2 - k_2 d k_1}{\kn (\kn -k_3)} 
\end{equation}
Using the spherical coordinates
\begin{align}
k_1 & = \kn \sin \theta \cos \varphi, & k_2 & = \kn \sin \theta \sin \varphi, & k_3 & =\kn \cos \theta
\end{align}
for $\kn>0$, $0< \theta \leq \pi$ and $0 \leq \varphi < 2 \pi$, one gets
\begin{equation}
\label{g1res}
\gmc{1} = \lambda \oint_C  2 \cos^2 \frac{\theta}{2}\, d\varphi
\end{equation}
In the special case when $\theta = \mathrm{const}$ and $\varphi$ changes from $0$ to $2 \pi$ the Berry phase (\ref{g1res}) reads
\begin{equation}
\gmc{1} = \lambda 4 \pi \cos^2 \frac{\theta}{2}\ = \lambda 2 \pi (\cos \theta +1)
\end{equation}
\end{exmpl}

\begin{exmpl}
\label{ex2}
Here we assume that $\Omega = \rtt \setminus \{ (0,0,k_3) \in \mathbb{R}^3 : k_3 < 0\} = \mathbb{R}^3 \setminus \{ (0,0,k_3) \in \mathbb{R}^3 : k_3 \in \mathbb{R}^1 \}$, and $a = a \ofk$ and $b= b \ofk$ are given by (\ref{abexmpl}). The Berry phase is now
\begin{multline}
\label{bpex2}
\gmc{2} = \lambda \oint_C \left( \frac{k_1 dk_2 -  k_2 dk_1}{\kn (\kn -k_3)} - d \varphi \right) = \lambda \oint_C (2 \cos^2 \frac{\theta}{2}-1)d\varphi \\
= \lambda \oint_C \cos{\theta} d\varphi = \gmc{1} - \lambda \oint_C d\varphi
\end{multline}
where $\gmc{1}$ is given by (\ref{g1res}). Therefore
\begin{equation}
\exp \Big( i \gmc{1}\Big) = \exp \Big( i \gmc{2}\Big)
\end{equation}
If $\theta = \mathrm{const}$ and $\varphi$ changes from $0$ to $2 \pi$ the Berry phase (\ref{bpex2}) is (see \cite{hawton,bialynicki})
\begin{equation}
\gmc{2} = \lambda 2 \pi \cos \theta
\end{equation}
\end{exmpl}

\begin{exmpl}
The domain $\Omega$ is as in Example \ref{ex2}; the functions $a$ and $b$ are given by (\ref{abexmpl2}). Now one quickly gets
\begin{equation}
\gmc{3} = \lambda \oint_C 2 (\cos^2 \frac{\theta}{2} -1)d \varphi = \gmc{2} - \lambda \oint_C d \varphi
\end{equation}
An important conclusion is that the phase factor $\exp \left( i \gmc{ } \right)$ is the same in all three examples
\begin{equation}
\exp \Big( i \gmc{1}\Big) = \exp \Big( i \gmc{2}\Big) = \exp \Big( i \gmc{3}\Big)
\end{equation}
Finally, when $\theta = \mathrm{const}$ and $\varphi$ changes from $0$ to $2 \pi$ (i.e.\ the closed loop goes around the $k_3$-axis) one has (see \cite{hawton3})
\begin{equation}
\gamma_3 \left[ C \right] = \lambda 2 \pi (\cos \theta -1) = -\lambda 4 \pi \sin^2 \frac{\theta}{2}
\end{equation}
\end{exmpl}

\section{Summary}
In the paper we have shown that the Hawton position operator for the photon with commuting components can be easily derived from the assumptions which are formulated in a natural manner within differential geometry language. We were able to find the general photon position operator satisfying those assumptions. 
Our approach enables one to find the eigenfunctions of the photon position operator in an easy way. These eigenfunctions in $\vec{x}$ representation are not spherically symmetric, which is in accordance with \cite{newton}. Of course, the spherical asymmetry is evident from the fact that the Hawton photon position operator depends on spin operator. We still cannot understand in all detail the properties of the eigenfunctions found in our paper but we are going to consider this soon. Another problem seems also to be interesting and is worth considering. In the previous work \cite{przanowski} devoted to the Weyl-Wigner-Moyal formalism of photon we concluded that in the phase space formulation for any quantum relativistic particle ``\ldots the problems with interpretation of the vector $\vec{x}$ are to be expected since for relativistic particles the operator $\hat{\vec{x}}$ does not represent the position observable \ldots ''. Therefore, an interesting question is if one can reformulate the Weyl-Wigner-Moyal formalism for the photon in such a way that instead of the operators $(\hat{\vec{p}},\hat{\vec{x}})$ the operators $(\hat{\vec{p}},\hat{\vec{X}})$ are applied. We are going to consider this question in the next work.

\section{Acknowledgments}
The work of F.~J.~T. was partially supported by SNI-México, COFAA-IPN and by SIP-IPN grants 20201186 and 20210759.

\end{document}